%% file: Bayesian_Optimization.tex
\documentclass[12pt]{article}
\input{Preamble}
\usepackage{authblk, nomencl, graphicx, commath, enumitem}
\usepackage[commentColor=black,noEnd=false,indLines=true]{algpseudocodex}

\usepackage{booktabs}
\usepackage{tikz,amsmath}
\usetikzlibrary{shapes,arrows}
\tikzset{
  font={\fontsize{10pt}{12}\selectfont}}
\makenomenclature
\usepackage{etoolbox}
\renewcommand\nomgroup[1]{%
  \item[
  \ifstrequal{#1}{A}{A. Abbreviations}{%
  \ifstrequal{#1}{B}{B. Capacity Expansion Problem}{%
  \ifstrequal{#1}{C}{C. Representative Period Clustering}{%
  \ifstrequal{#1}{D}{D. Bayesian optimization}{}}}}%
]}

\def\BibTeX{{\rm B\kern-.05em{\sc i\kern-.025em b}\kern-.08em
    T\kern-.1667em\lower.7ex\hbox{E}\kern-.125emX}}
\usepackage{balance}
\begin{document}
\input{tikz_set}

\begin{titlepage}
\title{Learning-assisted Stochastic Capacity Expansion Planning: A Bayesian Optimization Approach}
\author{Aron Brenner \and Rahman Khorramfar \and Dharik Mallapragada \and Saurabh Amin}

\date{\today}
\maketitle

\begin{abstract}
\noindent
Solving large-scale capacity expansion problems (CEPs) is central to cost-effective decarbonization of regional-scale energy systems. To ensure the intended outcomes of CEPs, modeling uncertainty due to weather-dependent variable renewable energy (VRE) supply and energy demand becomes crucially important. However, the resulting stochastic optimization models are often less computationally tractable than their deterministic counterparts. Here, we propose a learning-assisted approximate solution method to tractably solve two-stage stochastic CEPs. Our method identifies low-cost planning decisions by constructing and solving a sequence of tractable temporally aggregated surrogate problems. We adopt a Bayesian optimization approach to searching the space of time series aggregation hyperparameters and compute approximate solutions that minimize costs on a validation set of supply-demand projections. Importantly, we evaluate solved planning outcomes on a held-out set of test projections. We apply our approach to generation and transmission expansion planning for a joint power-gas system spanning New England. We show that our approach yields an estimated cost savings of up to 3.8\% in comparison to benchmark time series aggregation approaches.
\bigskip
\end{abstract}

\setcounter{page}{0}
\thispagestyle{empty}
\end{titlepage}

\pagebreak
\doublespacing

\section{Introduction}


\subsection{Motivation}
The transition from current fossil fuel-dominated energy systems to deeply decarbonized ones requires coordinated infrastructure planning and operations while accounting for uncertainties in key operational parameters that capture availability of weather-dependent variable renewable energy (VRE) supply and energy demand. Increasing penetration of VREs -- and increasing electrification of end-uses -- contribute to these uncertainties, with significant implications for energy infrastructure planning \cite{RoaldEtal2023Survey,ZhaoEtal2017}. Additionally, interdependencies between electric power and natural gas (NG) systems, as primary energy vectors, are intensified owing to the shifting role of NG-based generation to compensate for the intermittent nature of VREs, substitution of gas with electricity in end-uses (e.g. heating), and emerging technologies such as NG-based generation with carbon capture and storage \cite{KhorramfarEtal-AE2024}.

Capacity expansion problems (CEPs) form a crucial part of the energy systems planning toolkit as they guide infrastructure investment needed to meet future demand and decarbonization targets. Here, we focus on CEPs for low-carbon energy systems that incorporate multiple features. First, we account for the stochastic and weather-sensitive nature of key operational parameters in CEP models such as energy demand and VRE supply potential. Second, we recognize the importance of embedding operational dynamics such as ramping and storage at an hourly temporal fidelity. This becomes crucial as primary energy supply shifts to VRE generation and demand becomes more flexible \cite{KlatzerEtal2022, FarrokhifarEtal2020}. Finally, we consider the underlying network structure of energy vectors and their associated operating constraints in our CEP model to avoid underestimation of system cost \cite{FrysztackiEtal2021}. 

While including these aforementioned features allows us to more realistically capture system operations in CEP formulations, the resulting model is rendered computationally intractable due to its large-scale and mixed-integer nature. Even for simpler CEPs that omit these features, the literature on energy systems planning makes simplifying assumptions to trade-off model fidelity for computational tractability \cite{KotzurEtal2021,HelistoEtal2019}. Most commonly, planners resort to \textit{time series aggregation}, in which the CEP model is solved for a \textit{representative operational period} that (hopefully) spans a wide range of supply and demand patterns observed over the full operating horizon. Solving the resulting reduced-order CEP alleviates computational burden and can yield solutions that are comparable or similar to the solution of the CEP over the full planning horizon \cite{brenner2023, Teichgraeber2022Survey}. Unfortunately, these reduced-order CEPs do not account for the stochastic nature of input data -- particularly VRE resource potential and energy demand -- and thus have limited value from a planning perspective where cost effectiveness and constraint satisfaction under future scenarios are important criteria. On the other hand, solving full-scale stochastic CEPs with multiple scenarios of VRE and demand realization becomes computationally intractable with only a handful of scenarios, particularly when considering discrete investment decisions. This challenge necessitates a new approach that integrates the stochastic nature of VRE and demand parameters in CEPs while remaining computationally scalable.

\subsection{Time Series Aggregation}
Time series aggregation for energy systems has a rich literature \cite{Teichgraeber2022Survey,HoffmannEtal2020-survey}. In the context of planning problems, researchers select representative periods based on electric power demand and/or capacity factors (CFs) for VREs \cite{TeichgraeberBrandt2019,li2022,SunEtal2019, brenner2023}. To automatically select representative periods from this supply-demand data, one solves a \textit{representative period clustering} (RPC) problem using well-known clustering algorithms such as k-means \cite{MallapragadaEtal2018,TeichgraeberBrandt2019,li2022}, k-medoids \cite{TeichgraeberBrandt2019,li2022,brenner2023}, or hierarchical clustering \cite{Pfenninger2017,SunEtal2019}.

Although clustering approaches to time series aggregation partially address the computational challenge in solving CEPs at scale, they face two major shortcomings. First, hyperparameters of the RPC problem are often selected in an ad hoc manner; these include the number of representative periods and the relative weights of various supply and demand features in the distance metric used for clustering. This precludes exploring a range of potential CEP instances that can differ in terms of inputs such as network topology, resource availability, and demand patterns. Additionally, hyperparameters selected for one CEP instance may not yield effective planning outcomes for other instances. Second, despite the well-known fact that weather variation impacts both supply and demand projections  \cite{MallapragadaEtal2018}, typical approaches to time series aggregation largely ignore this uncertainty in selecting representative periods. Instead, the performance of clustering methods is simply evaluated by quantifying their ability to reproduce the CEP investment decisions using a \textit{single} supply-demand projection \cite{li2022, TeichgraeberEtal2020,PinedaMorales2018}. This yields planning outcomes that are tailored to a specific projection of supply-demand data but might be hugely suboptimal for supply-demand patterns that might be realized for weather-years in the planning horizon. In effect, existing clustering methods applied to CEPs may produce investment outcomes that are not optimal under inter-annual variability. As weather-dependent VRE supply is expected to grow with decarbonization efforts, there is growing interest from researchers \cite{oree2017, koltsaklis2018state} and system operators \cite{ISONE2023} to identify CEP investment outcomes that \textit{meet demands at low cost under uncertain VRE supply and energy load profiles}.

\begin{figure}[htbp]
    \centering
    \includegraphics[width=0.6\textwidth]{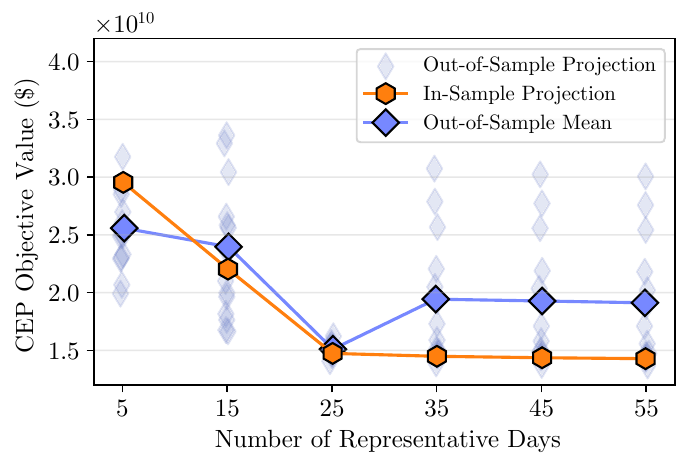}
    \caption{Effect of varying the number of representative days on the CEP objective evaluated over 14 out-of-sample supply-demand projections using a single projection solution for a joint power-gas CEP (Sec.~\ref{sec:experiments}).}
    \label{fig:intro}
\end{figure}

Fig.~\ref{fig:intro} highlights the implications of these shortcomings in the case of a CEP for joint power-NG system planning \cite{KhorramfarEtal-AE2024}. We observe that investment decisions that minimize costs for a single supply-demand projection can yield much higher operational costs for different realization of demands and VRE availabilities (as seen by out-of-sample projections shown). We also notice that average costs incurred over 14 \textit{out-of-sample} projections (i.e., supply-demand projections that are not used to instantiate the model) do not decrease monotonically with the number of representative days and are instead minimized at $25$ representative days. When using more than 25 representative days, high out-of-sample costs may result from ``overfitting'' of investment decisions to the single projection used to instantiate the \textit{surrogate} (i.e., reduced-order) CEP. This phenomenon necessitates careful tuning of RPC hyperparameters with consideration of \textit{multiple} supply-demand projections and systematic \textit{out-of-sample cost evaluations}. Our work addresses this challenge using a learning-assisted approach to time series aggregation.

\subsection{Contributions}
We consider two-stage stochastic CEP for coordinated planning of multiple energy vectors. We develop a learning-assisted approach to compute investment decisions that yield the lowest combined investment and operational cost in expectation over multiple year-long projections of energy demand and VRE availability (Sec.~\ref{sec:problem}). Rather than solve an extensive form stochastic program, which is intractable due to the large-scale and combinatorial nature of the problem, our approach designs and solves a surrogate CEP instantiated using a smaller set of representative periods. Note, however, that there can be many potential surrogate CEPs, each yielding different planning outcomes. Our key contribution is to leverage a Bayesian optimization (BO) approach to efficiently search the space of RPC hyperparameters. We demonstrate that these automatically tuned hyperparameters help identify planning outcomes with the lowest expected total cost (investment plus operational) over \textit{out-of-sample} supply-demand projections  (Sec.~\ref{sec:methods}).

\begin{figure*}[htbp]
    \centering
    \begin{subfigure}[t]{0.55\textwidth}
        \centering
        \begin{subfigure}[t]{0.25\textwidth}
            \centering
            \begin{tikzpicture}[auto, node distance=1.6cm, >=latex', scale=0.7, every node/.style={transform shape}]
            \input{Tikz_Figures/learning-to-configure}
            \end{tikzpicture}
            \caption*{Training}
        \end{subfigure}
        \hfill
        \begin{subfigure}[t]{0.3\textwidth}
            \centering
            \begin{tikzpicture}[auto, node distance=1.6cm, >=latex', scale=0.7, every node/.style={transform shape}]
            \input{Tikz_Figures/learning-to-configure-deploy}
            \end{tikzpicture}
            \caption*{Deployment}
        \end{subfigure}
        \caption{Learning-assisted heuristics}
    \end{subfigure}
    \hfill
    \begin{subfigure}[t]{0.3\textwidth}
        \begin{subfigure}[t]{0.25\textwidth}
            \centering
            \begin{tikzpicture}[auto, node distance=1.6cm, >=latex', scale=0.7, every node/.style={transform shape}]
            \input{Tikz_Figures/BO}
            \end{tikzpicture}
            \caption*{}
        \end{subfigure}
        \caption{Our BO-assisted approach}
    \end{subfigure}
    \caption{Conceptual difference between offline learning-assisted heuristic approaches in training and deployment (a) and our proposed BO-assisted approach (b). Here, $\theta$ denotes the heuristic hyperparameters, $x$ denotes the decision variables of the optimization task, and ML denotes the machine learning module (i.e., BO in our approach). Dashed lines indicate the flow of decision costs to ML model estimation.}
    \label{fig:conceptual-comparison}
\end{figure*}

Our approach exploits two properties shared by many CEPs for energy systems planning: (1) while optimizing investment decisions for a full planning horizon is difficult (i.e., mixed-integer), evaluating the operational cost incurred by a fixed set of investment decisions often reduces to solving a linear program for each projection; (2) investment decisions obtained from solving a surrogate problem instantiated over a small number of representative periods can yield more robust planning outcomes (in terms of better out-of-sample performance) with less computational effort in comparison to solving a larger surrogate problem. This second property has not been explored in the existing literature. We show how these properties can be systematically leveraged to tractably solve two-stage stochastic CEPs.

Our BO-assisted approach can be viewed as a ``learning to configure'' approach \cite{bengio2021} in that it learns to select hyperparameters specifying the time series aggregation heuristic for surrogate model construction. In this regard, our approach is distinct from recent works that use offline learning to improve optimization heuristics \cite{han2023, prat2023} as it can be deployed \textit{without pre-training} on a large number of problem instances (Fig.~\ref{fig:conceptual-comparison}). Specifically, our approach quickly identifies low-cost investment decisions by learning to identify promising hyperparameter configurations for a \textit{single problem instance} through an iterative process of instantiating surrogate problems, solving them, and evaluating the resulting investment decisions over a range of validation supply-demand projections. We demonstrate the effectiveness of this approach for joint power-gas generation and transmission expansion planning for New England with 20 yearly supply-demand projections (Sec.~\ref{sec:experiments}).

\section{Problem Formulation}\label{sec:problem}
\subsection{Capacity Expansion Problem}\label{sec:cep}
We consider a stochastic generalization of two-stage CEPs that commonly arise in energy systems planning:
\begin{subequations}\label{cep-original}
\begin{align}
    [\text{CEP}] \quad \min_{{x}} \quad & {c}^\top {x} + \mathbb{E}_{\omega\in \Omega} [Q({x}, \omega)] \label{s1-obj}\\
    \text{s.t.} \quad & {A}{x} = {b} \label{s1-c1}.
\end{align}
\end{subequations}
The first-stage variable ${x}$ takes integer and continuous values and denotes the investment or decommissioning decisions that are realized before the planning horizon. These describe locations and investment levels for various plant types (e.g., solar panels, wind turbines, gas-fired plants), transportation (i.e., transmission lines for power systems, pipelines for NG systems), and storage facilities. Constraints~\eqref{s1-c1} imposes first-stage constraints such as budget and number of operational assets. The objective function~\eqref{s1-obj} minimizes the combined investment and operational cost over a scenario set, $\Omega$.

\begin{figure}[htbp]
    \centering
    \includegraphics[width=0.6\textwidth]{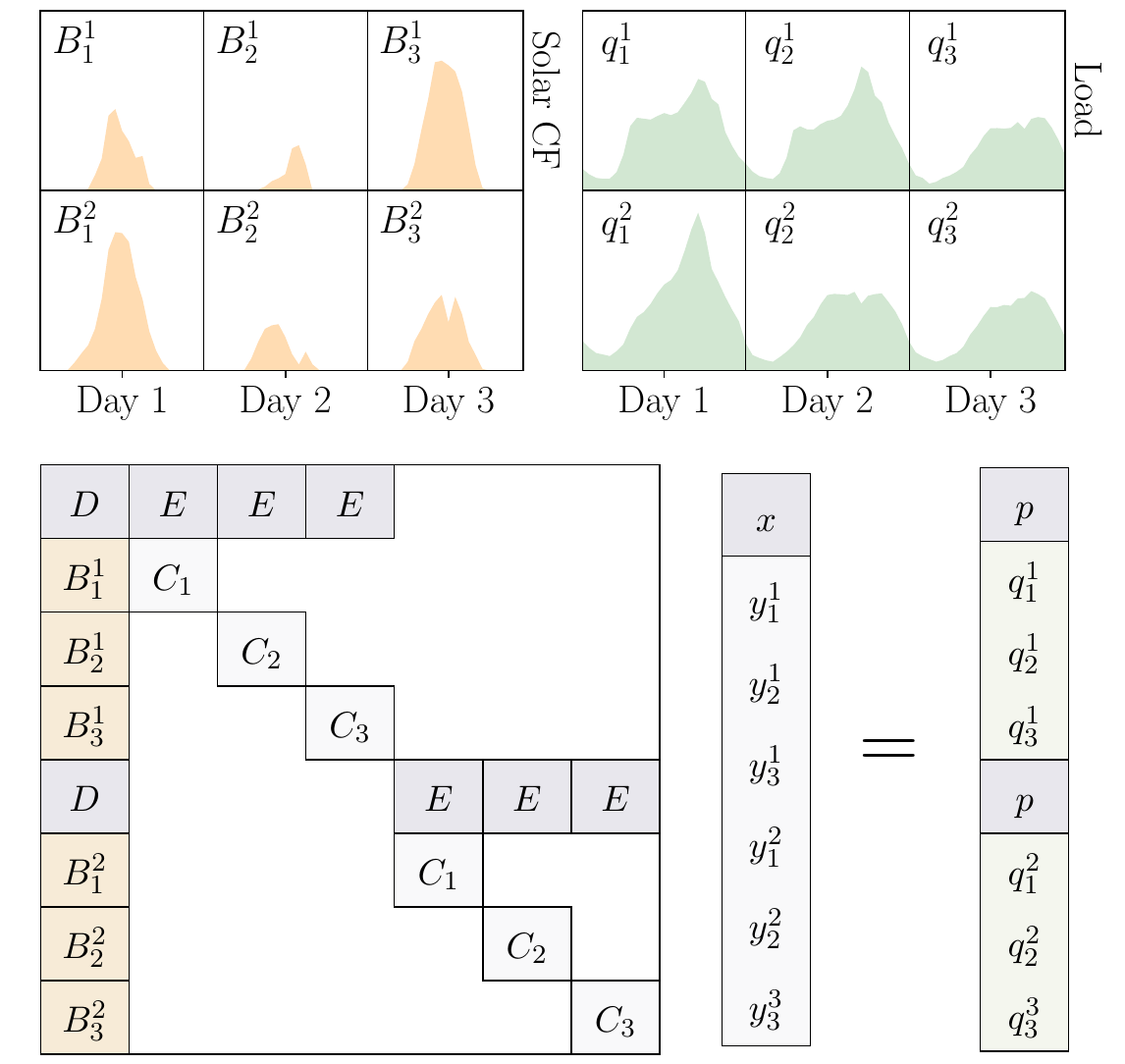}
    \caption{Supply-demand projections \cite{KhorramfarEtal-NE2024} (top) and constraints (bottom) for an illustrative stochastic capacity expansion problem with two projections and three periods (i.e., days).}
    \label{fig:block_matrix}
\end{figure}

For a feasible first-stage solution $x$, the value function of the recourse problem for each year-long stochastic scenario (i.e., supply-demand projection) $\omega \in \Omega$ is given by
\begin{subequations} \label{model:s2}
\begin{align}
    Q(x,\omega) \ = \ \min_{{y}} \quad & \sum_{t \in \mathcal{T}} {d}_t^\top {y}_t \label{s2-obj}\\
    \text{s.t.} \quad & {B}_t^\omega {x} + {C}_t{y}_{t} = {q}_t^\omega & t \in \mathcal{T} \label{s2-c1} \\
    & {D} {x} + \sum_{t\in\mathcal{T}} E {y}_t = {p} \label{s2-c2},
\end{align}
\end{subequations}
where ${y}_t$ is a continuous variable that denotes second-stage decisions in period $t$. Without loss of generality, we assume that each $t\in \mathcal{T}$ corresponds to a single day of the planning period. The objective function~\eqref{s2-obj} minimizes operational costs, which can include variable generation, load shedding penalty and fuel costs. The constraints~\eqref{s2-c1} link capacity expansion decisions in the first stage to operational decisions in the second stage. In particular, these constraints ensure energy balance of demands ${q}_t^\omega$, load shedding ${y}_t$, and generation constrained by first-stage decision variables and renewable capacity factors ${B}_t^\omega$. The second set of constraints~\eqref{s2-c2} links operational decisions across periods such as ramping and storage for adjacent hours and cross-sectoral emissions limits throughout the entire planning horizon. Here, all inequality constraints are formulated as equality constraints using slack variables. Fig.~\ref{fig:block_matrix} illustrates how supply and demand projections enter into the problem to form the constraint matrix.

We assume that~\eqref{model:s2} is feasible for any first-stage decisions, $x$, i.e., load shedding is unconstrained and incurred as a cost. Additionally, we note that each period, $t\in\mathcal{T}$, does not necessarily correspond to the planning resolution of energy systems. For instance, while hourly time resolution is usually used for electric power planning, daily operating resolution can be more suitable for gas system planning \cite{KhorramfarEtal-AE2024}. Following \cite{li2022, poncelet2016, almaimouni2018}, we assume representative periods to denote daily time spans. 

\subsection{Surrogate Capacity Expansion Problem} \label{sec:surrogate}
In practice, the number of second-stage decision variables and constraints, $\mathcal{O}(|\Omega||\mathcal{T}|)$, is orders of magnitude larger than those corresponding to the first stage. This renders an intractable formulation even for CEPs that consider a single scenario (on the order of $10^6$ variables and constraints in our computational study with an hourly resolution). As a result, scenario-wise decomposition methods cannot be applied to solve~\eqref{cep-original} quickly. Furthermore, applying a day-by-day decomposition approach requires relaxing inter-day linking constraints \cite{nasri2015network}. Consequently, we must consider approximate solution algorithms for~\eqref{cep-original}. Such  an approach instead solves a surrogate model -- a reduced-order formulation of  $\eqref{cep-original}$ instantiated on a representative day set, $\mathcal{T}_R \subset \mathcal{T}$, with a reduced number of variables and constraints, $\mathcal{O}(|\Omega||\mathcal{T}_R|)$.

To formulate our surrogate model, we first construct the unordered sample set $\mathcal{S}$, which contains daily supply-demand profiles for the entire planning horizon for all scenarios. Specifically, for a planning horizon of one year, $\mathcal{S}$ contains $365 \times |\Omega|$ samples. This sample set $\mathcal{S}$ is distinguished from the notion of a scenario set as a scenario $\omega \in \Omega$, describes a \textit{year-long} supply-demand projection. On the other hand, the sample set describes the unordered collection of \textit{day-long} supply-demand projections taken from across all scenarios.

Next, we select $\theta_R$ days from $\mathcal{S}$ and construct the set of representative days $\mathcal{T}_R$. Additionally, we assign a weight $w_t\in \mathcal{W}$ to each representative day in $\mathcal{T}_R$ such that $\sum_{t\in\mathcal{T}_R}w_t = |\mathcal{T}|$. The resulting surrogate problem can be formulated as the single-scenario CEP,
\begin{subequations}\label{model:REP}
\begin{align}
    [\text{CEP}_\text{S}(\mathcal{T}_R, \mathcal{W})] \quad \min_{{x}, {y}} \quad & {c}^\top {x} + \sum_{t \in \mathcal{T}_R} w_t {d}_t^\top {y}_t\\
    \text{s.t.} \quad & {A}{x} = {b} \\
    & {B}_t^\omega {x} + {C}_t{y}_{t} = {q}_t^\omega, \hspace{2mm} t \in \mathcal{T}_R \\
    & {D} {x} + \sum_{t\in\mathcal{T}_N} w_tE {y}_t = {p}.
\end{align}
\end{subequations}
We note that different orderings of days in $\mathcal{T}_R$ instantiate different surrogate problems. For simplicity, we assume that days in $\mathcal{T}_R$ are arranged in an increasing order of the scenario index and day of year.

\subsection{Representative Period Clustering Problem}\label{sec:rpc}
The choice of representative days is crucial to achieving a computationally light surrogate model that well-approximates~\eqref{cep-original}. In other words, the investment decisions $x$ resulting from $\text{CEP}_\text{S}(\mathcal{T}_R, \mathcal{W})$ should approximately minimize~\eqref{s1-obj}. Existing literature, including \cite{ZhangDharik2022rep_day, poncelet2016, SunEtal2019, TeichgraeberBrandt2019}, propose clustering-based approaches for identifying such representative periods. Here, we consider the following k-medoids clustering model, which we refer to as the representative period clustering (RPC) problem:
\begin{subequations}\label{eq:TSA}
\begin{align}
    [\text{RPC}(\theta)] \quad \min \quad & \sum_{i = 1}^{k} \sum_{t\in\mathcal{C}_i} D(m_i, t) \\
    \text{s.t.} \quad & \bigcup_{i=1}^k \mathcal{C}_i = \mathcal{S} \label{eq:rpc_objective}\\
    & \mathcal{C}_i \cap \mathcal{C}_j = \emptyset \hspace{10mm} \forall i=[\theta_R], j\neq i \\
    & (m_1,\dots,m_{\theta_E}) = \mathcal{T}_E \label{eq:extreme1} \\
    & \mathcal{C}_i = \{t\} \hspace{22mm} \forall t \in \mathcal{T}_E, \label{eq:extreme2}
\end{align}
\end{subequations}
where $\mathcal{C} \coloneqq (\mathcal{C}_1, \dots, \mathcal{C}_{k})$ denotes the representative day clusters with corresponding medoids $(m_1, \dots, m_{k})$, $D(m_i, t)$ denotes the ``distance'' between medoid $m_i$ and period $t$, and $\mathcal{T}_E \subset \mathcal{T}_R$ denotes a set of ``extreme days,'' or days exhibiting particularly hard-to-satisfy demands, with corresponding weights $w_i=1$ for all $i \in \mathcal{T}_E$. One can apply an out-of-the-box clustering algorithm \cite{wiley1990} to solve~\eqref{eq:TSA} and obtain the representative day set and weights respectively as $\mathcal{T}_R = (m_1,\dots,m_{\theta_R})$ and $\mathcal{W} = (\frac{1}{|\mathcal{S}|}|\mathcal{C}_1|, \dots, \frac{1}{|\mathcal{S}|}|\mathcal{C}_{\theta_R}|)$. Here, we focus on the central issue of selecting hyperparameters for the RPC problem.

Before proceeding, we note that~\eqref{eq:extreme1} ensures that the representative day set $\mathcal{T}_R$ contains $\theta_E$ ``extreme'' days, $\mathcal{T}_E$, or days with high energy demands. Additionally, constraint~\eqref{eq:extreme2} enforces that each extreme day forms a single-member cluster. Including extreme days in the surrogate model may compensate for the smoothing effect of clustering and result in a planning outcome that is more robust to peak demands. However, selecting the hyperparameter $\theta_E$ is not obvious a priori and again requires tuning to avoid over- or under-investment.

The distance function, $D(m_i, t)$ is another choice that greatly impacts the RPC solution, and consequently, planning outcomes. A natural choice is the Euclidean distance between supply-demand parameters associated with a pair of days,
\begin{align*}
    D(t, t') &= \theta_B \|{B}_t^\omega - {B}_{t'}^\omega\|_F + (1-\theta_B) \|{q}_t^\omega - {q}_{t'}^\omega\|_2,
\end{align*}
where $\|\cdot\|_F$ denotes the Frobenius norm. We consider $\theta_B \in [0,1]$ to be a hyperparameter that weighs the relative importance of demand (contained in ${q}_t^\omega$) and supply-side features such as renewable capacity factors (contained in ${B}_t^\omega$). Importantly, different choices of $\theta_B$ impact RPC outcomes by influencing the daily supply- or demand- patterns captured in the representative day set. Selecting $\theta_B$ is not obvious. More generally, more than two groups of parameters may need to be weighed when multiple VRE technologies and energy vectors are considered, which is the case for our computational study (Sec.~\ref{sec:experiments}).

\begin{figure}[h]
    \centering
    \includegraphics[width=0.6\textwidth]{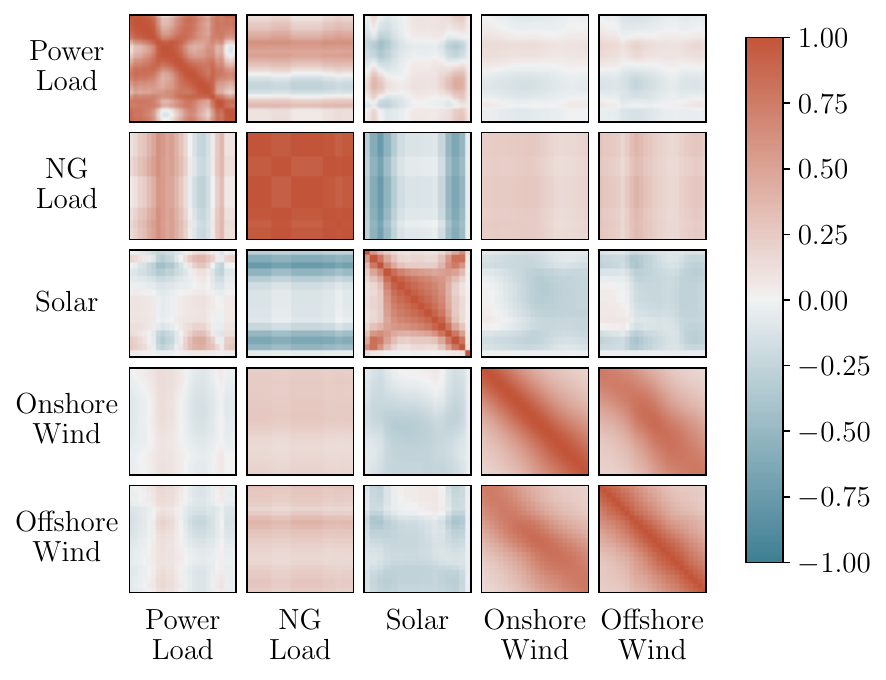}
    \caption{Pearson correlations between supply-demand parameter groups for our computational study (Sec.~\ref{sec:experiments}). Each row/column of pixels corresponds to one hour of the day (averaged over system nodes) except for NG, where pixels correspond to daily nodal loads. For each pair of parameter groups, pixel $(i,j)$ shows the correlation of parameter $i$ in the first group with parameter $j$ in the second group.}
    \label{fig:corr}
\end{figure}

Fig.~\ref{fig:corr} illustrates one challenge in choosing $\theta_B$ that results from correlations among supply-demand parameters. For example, clustering days according to, e.g., NG loads, will also capture a wide range of early morning/late evening solar availability due to highly negative correlations between NG demand and solar availability resulting from the time of year. However, wind availability patterns may not be proportionately represented in the surrogate problem due to minimal correlation with NG loads. In the presence of such correlations, an RPC problem that weighs all parameters equally may, in fact, fail to adequately capture variability of certain parameters and consequently yield suboptimal planning outcomes.

Essentially, RPC hyperparameters can singificantly impact planning outcomes obtained by solving $\text{CEP}_\text{S}$. Previous works do not directly deal with this issue and instead focus on evaluating outcomes of individual hyperparameter choices, namely $\theta_R$ \cite{TeichgraeberBrandt2019, MallapragadaEtal2018} and $\theta_E$ \cite{li2022}, for specific planning contexts/formulations. In particular, most works that cluster according to both supply- and demand-side parameters avoid applying any weighting and select $\theta_B$ implicitly \cite{li2022, MallapragadaEtal2018}. 

Here, we highlight another limitation of prior work, specifically the assumption that planning outcomes improve monotonically in $\theta_R$ \cite{TeichgraeberBrandt2019, SunEtal2019}. Consequently, their results suggest that the ideal choice of $\theta_R$ is a point at which planning objectives \cite{TeichgraeberBrandt2019, SunEtal2019} or clustering objectives \cite{AlmaimouniEtal2018} show diminishing returns with respect to computational runtime. In contrast, we find that this is not the case for out-of-sample evaluations, as illustrated in Fig.~\ref{fig:intro}. In particular, the ideal choices of $\theta_R$, $\theta_E$, and $\theta_B$ depend on the problem and the available data that can be used to instantiate the surrogate problem and evaluate resulting costs out-of-sample. The difficulty of navigating this hyperparameter space, which varies widely across problem settings, necessitates developing a systematic approach to \textit{jointly} tuning multiple hyperparameters in a cost-driven manner. Next, we describe and evaluate a BO-assisted approach for hyperparameter tuning based on resulting out-of-sample cost objectives.

\begin{figure*}[h]
    \centering
    \includegraphics[width=\textwidth]{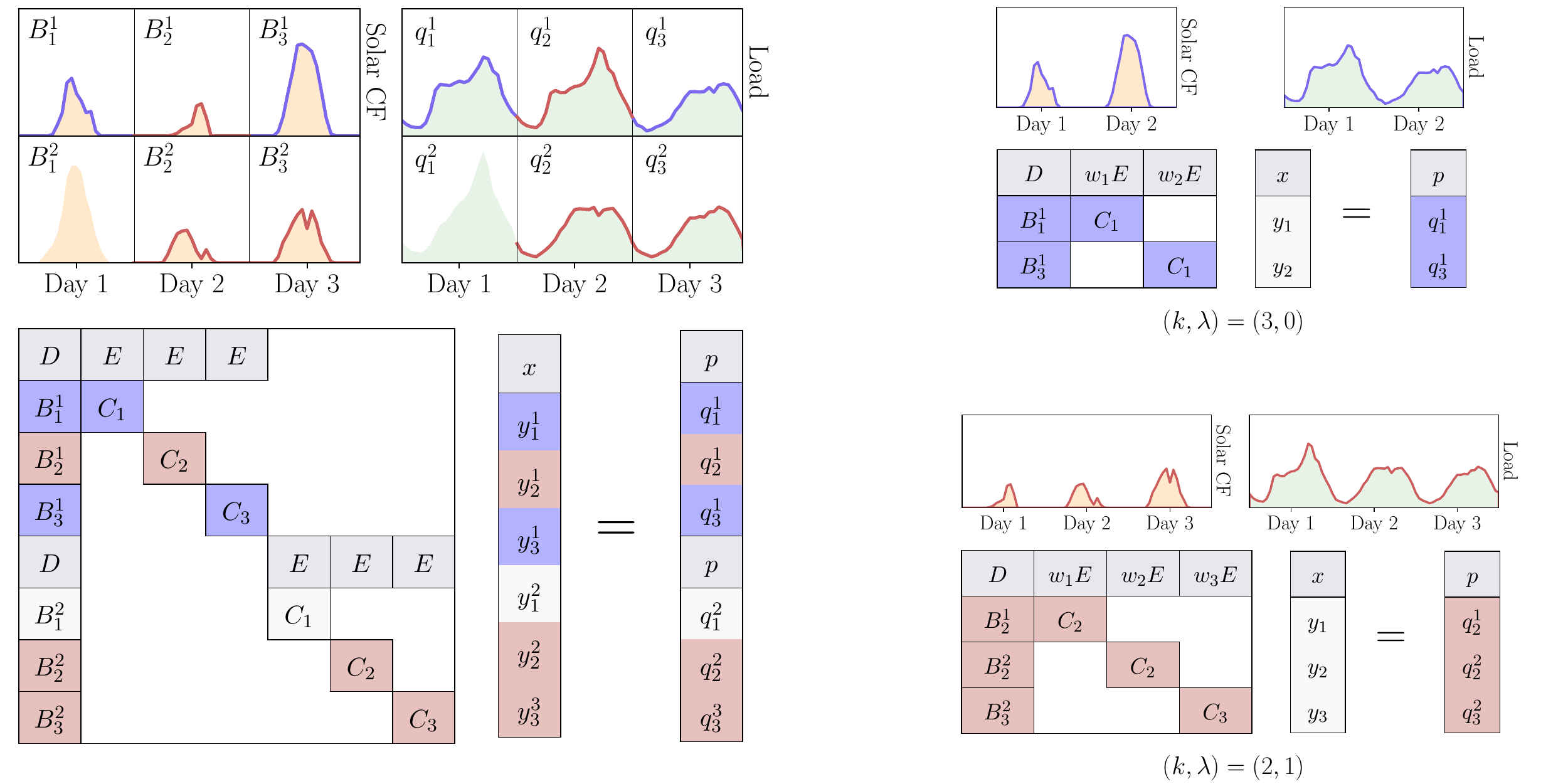}
    \caption{Surrogate problems are instantiated as ``single-scenario'' CEPs with a representative day set constructed using days sourced from different supply-demand projections. Different choices of RPC hyperparameters instantiate different surrogate problems. Setting $(k,\lambda) = (2,1)$, the resulting surrogate problem (top right) captures typical solar availability and nominal energy demand patterns. Setting $(k,\lambda) = (3,0)$, the resulting surrogate problem (bottom right) captures a wider range of load profiles but fails to capture days with high solar availability.}
    \label{fig:clustered_matrix}
\end{figure*}

 \begin{figure*}[hbtp]
    \centering
    \includegraphics[width=\textwidth]{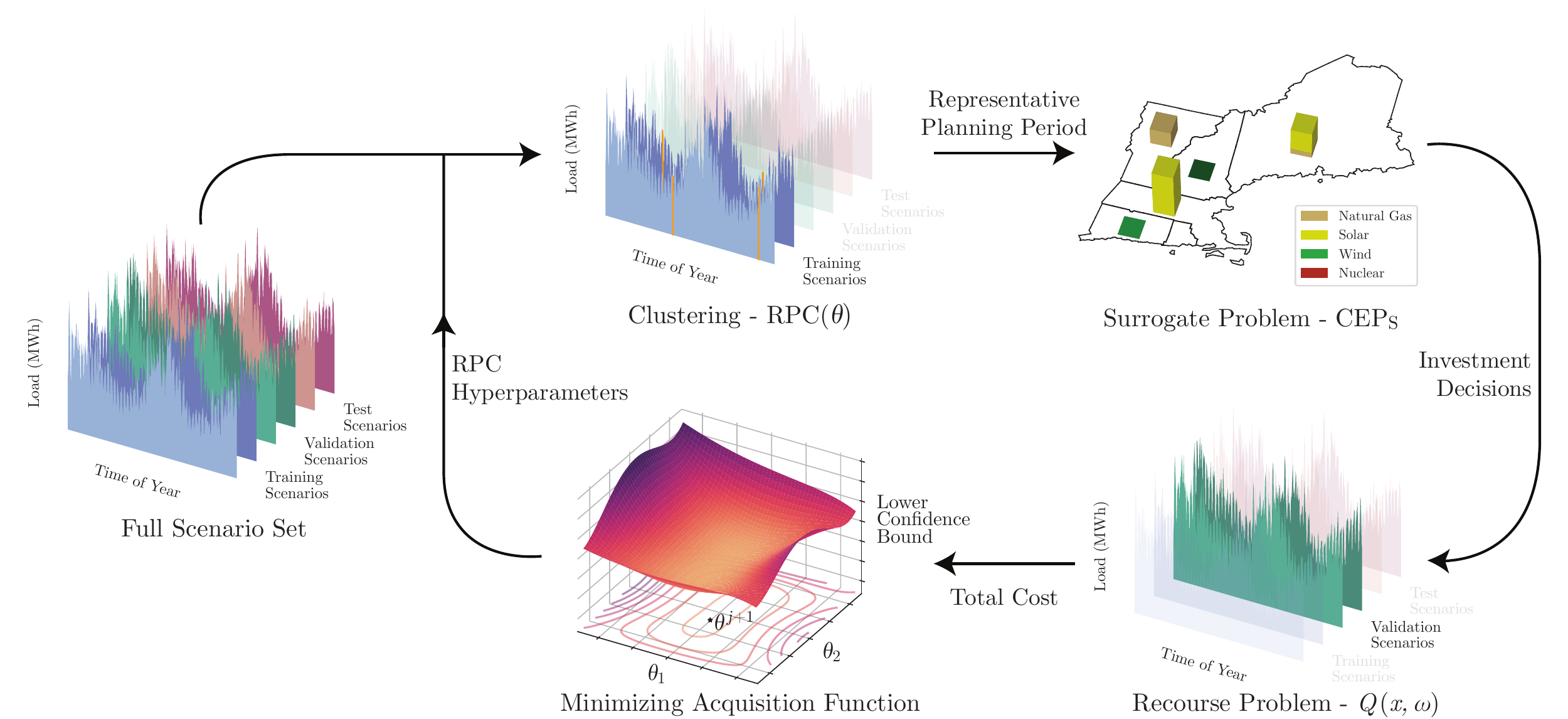}
    \caption{The proposed BO-assisted solution approach.}
    \label{fig:main_fig}
 \end{figure*}

\section{Solution Approach}\label{sec:methods}
To address this challenge systematically and for a range of planning contexts, we propose searching over a continuous space of RPC hyperparameters using Bayesian optimization (BO), a derivative-free search strategy for black-box functions that has been applied with success in hyperparameter tuning for ML applications \cite{snoek}. At a high level, our BO-assisted approach learns to minimize $f(\cdot)$, the function mapping RPC hyperparameters to CEP objective values, by performing a series of \textit{function evaluations}. As is the case in many BO applications, $f(\cdot)$ does not admit an analytic expression and can only be evaluated through a series of complex operations (i.e., solving optimization problems). In each iteration of our approach, a function estimate -- a Gaussian process (GP) regression model -- is re-estimated using the most recent function evaluation, and new candidate hyperparameters are identified according to their estimated potential for yielding a low CEP objective and improving the GP model estimate. Fig.~\ref{fig:main_fig} illustrates our solution approach and shows how hyperparameters $\theta$ are evaluated (Sec.~\ref{sec:function_evaluation}) and how subsequent candidate hyperparameters are selected (Sec.~\ref{sec:GP}) in each iteration.

\subsection{Function Evaluation}\label{sec:function_evaluation}
Before describing the function evaluation process, we discuss the importance of dividing scenarios into disjoint sets of training and validation to reduce ``look-ahead'' bias. This is an optimistic bias in the estimate of operational cost that occurs as a result of including days from the second stage in the representative day set. In other words, if the scenario sets are not separated during training, the BO approach might learn to include specific days from the validation set that minimize validation costs. Consequently, the corresponding operational cost estimate will be lower, on average, than the actual operational cost incurred out-of-sample (assuming projections are drawn i.i.d.) as the representative day set has been ``overfit'' to the sample set (see Fig.~\ref{fig:intro}). 
Accordingly, we split the scenario set $\Omega$ into training, validation, and test sets, $\Omega^\text{train}, \Omega^\text{val}, \Omega^\text{test}$ and evaluate costs over $\Omega^\text{val}$ in order to obtain a less biased estimate of costs incurred on $\Omega^\text{test}$ (i.e., out-of-sample).

Denoting the number of RPC hyperparameters by $k$, we define $\Theta \subset \mathbb{R}^k$ to be the bounded space of candidate RPC hyperparameters. We assume $\Theta$ to be continuous, which we generalize to integer-valued hyperparameters with rounding. In each iteration of the search, we evaluate $f(\theta)$ for some candidate hyperparameters $\theta \in \Theta$. For example, if we want to search over the space of $10 \leq \theta_R \leq 20$ and $1 \leq \theta_E \leq 10$, we define $\Theta \coloneqq [10,20] \times [1,10]$ and in each iteration evaluate some candidate $\theta = (\theta_R, \theta_E)$. This function evaluation requires three steps.

1) \textbf{Clustering the sample set.} We solve~\eqref{eq:TSA} using a sample set constructed from $\Omega^\text{train}$ to identify a set of representative days (i.e., cluster medoids) and their corresponding weights, $\mathcal{T}_R$ and $\mathcal{W}$ respectively.

2) \textbf{Obtaining investments.} We then solve~\eqref{model:REP} for the given representative days to obtain investment decisions, ${x}$, and the corresponding investment cost, ${c}^\top {x}$.

3) \textbf{Evaluating operational costs.} Given the investment decisions $x$ from the second step, we solve the resulting recourse problem~\eqref{model:s2} for the full planning horizon, $\mathcal{T}$, for all scenarios in $\Omega^\text{val}$. This yields a set of operational decisions and corresponding operational costs for all validation scenarios. Averaging these values gives $f(\theta) = {c}^\top {x} + \frac{1}{|\Omega^\text{val}|}\sum_{\omega \in \Omega^\text{val}} \sum_{t\in\mathcal{T}} d_t^\top y_t^\omega$, an upper bound (i.e., feasible solution) for~\eqref{s1-obj} instantiated on $\Omega^\text{val}$.

Regarding computational runtime, solutions for $\text{RPC}(\theta)$ can be obtained quickly with approximate algorithms such as PAM \cite{wiley1990}. The computational burden in Step (3) is also likely to be low as it requires solving $|\Omega^\text{val}|$ recourse problems, which are linear programs in our case. Additionally, Step (3) can easily be parallelized to accelerate the process. In most cases, Step (2) of the function evaluation will require the greatest runtime as it involves solving a mixed-integer linear program~\eqref{model:REP} with $\mathcal{O}(|\mathcal{T}_R|)$ second-stage variables and constraints. Consequently, the worst-case runtime of one function evaluation depends on the maximum value that $\theta_R$ (i.e., the number of representative days) takes in $\Theta$.

\subsection{Selecting Candidate Hyperparameters}\label{sec:GP}
Following each function evaluation, the procedure identifies the next candidate hyperparameters in three steps. Let us suppose that we have already evaluated $j$ hyperparameter settings, $\theta^1, \dots, \theta^j$, which has returned function values $f(\theta^1), \dots, f(\theta^j)$.

1) \textbf{Instantiating a prior.} We instantiate a prior predictive covariance function, or \textit{kernel function}, $\Sigma: \Theta \times \Theta \to \mathbb{R}$, which yields the GP prior distribution
\begin{align}\label{eq:GP_prior}
    f^{1:j} \sim \mathcal{N}(0, \Sigma({\theta}^{1:j},{\theta}^{1:j})),
\end{align}
where ${\theta}^{1:j}$ denotes $\theta^1, \dots, \theta^j$ and the kernel function is defined as:
\begin{align*}
    \Sigma({\theta}, {\theta}') = \frac{2^{-3/2}}{\Gamma(5/2)} (\sqrt{5}\|{\theta} - {\theta}'\|_2)^{5/2} K_{5/2} (\sqrt{5}\|{\theta} - {\theta}'\|_2),
 \end{align*}
where $\Gamma(\cdot)$ and $K_{5/2}(\cdot)$ are the gamma function and modified Bessel function respectively \cite{frazier2018}. This is a common choice of kernel function \cite{frazier2018, snoek} and is the default setting in \textsc{BoTorch} \cite{botorch}, the package that we used in our experiments. Note that in~\eqref{eq:GP_prior} we assume $f^{1:j}$ to have zero mean, which we enforce in practice through standardization.

2) \textbf{Performing a Bayesian update}. We then estimate the GP posterior conditional distribution of CEP validation costs over the hyperparameter space by applying a Bayesian update. This posterior is given by
\begin{align*}
    f \mid \theta, \theta^{1:j}, f^{1:j} \sim \mathcal{N}(\mu(\theta), \sigma^2(\theta)),
\end{align*}
where
\begin{align}
    \mu(\theta) &= \Sigma(\theta, \theta^{1:j}) \Sigma(\theta^{1:j}, \theta^{1:j})^{-1} f^{1:j} \label{eq:GP_update1}\\
    \sigma^2(\theta) &= \Sigma(\theta, \theta) - \Sigma(\theta, \theta^{1:j}) \Sigma(\theta^{1:j}, \theta^{1:j})^{-1} \Sigma(\theta^{1:j}, \theta) \label{eq:GP_update2}
\end{align}
are the posterior predictive mean and variance functions respectively \cite{rasmussen2006}.

3) \textbf{Minimizing the acquisition function.} To identify promising candidate hyperparameters, we must define an \textit{acquisition function} that quantifies potential for evaluation as a function of the estimated GP posterior. Accordingly, we consider the \textit{lower confidence bound} (LCB) acquisition function
\begin{align}\label{eq:LCB}
    [\text{LCB}] \qquad & a_\text{LCB}(\theta) = \mu({\theta}) - \beta \sigma^2 ({\theta}),
\end{align}
and in each iteration select ${\theta}^{j+1}$ that minimizes $a_\text{LCB}(\theta)$. Here, $\beta$ is a GP hyperparameter (not to be confused with hyperparameters of the RPC problem) that tunes the exploration-exploitation tradeoff of selecting candidates with low predicted means as opposed to those with high predicted variances \cite{snoek}. We note that the second stage will always be feasible for any choice of investment decisions, $x$, and by extension, RPC hyperparameters, $\theta$. Consequently,~\eqref{eq:LCB} is unconstrained and can be solved quickly using quasi-Newton methods such as BFGS \cite{nocedal1999}, which are implemented as part of standard Bayesian optimization packages.



Most implementations of BO approaches recommend initializing the GP model by evaluating a set of $N_0$ points, ${\theta}^{1}, \dots, {\theta}^{N_0}$ sampled from some distribution over the search space $\Theta$ \cite{frazier2018, snoek}. Initializing the GP with a larger random sample improves the estimate of the GP but increases the number of function evaluations required before ``higher potential'' candidates can be evaluated. Properly normalizing $\theta^1, \dots, \theta^j$ and $f(\theta^1), \dots, f(\theta^j)$ before estimating the GP also has a significant impact on the performance of BO in practice. Following \textsc{BoTorch} specifications \cite{botorch}, we rescale $\Theta$ to $[0,1]^m$ before estimation. Additionally, we standardize the function evaluations so that $f(\theta^1), \dots, f(\theta^j)$ has zero mean and unit variance in each iteration before estimating the GP.

\section{Computational Study}\label{sec:experiments}
In this section, we evaluate the ability of our proposed BO-assisted approach to identify low-cost planning outcomes by searching the low-dimensional space of RPC hyperparameters for a stochastic variant of the Joint Power and Natural Gas (JPoNG) model \cite{KhorramfarEtal-AE2024}. The JPoNG model evaluates the cost-optimal planning of power and NG infrastructure with a resolved representation of spatial, temporal, and technological system constraints. The model structure follows that of a two-stage stochastic CEP that minimizes annualized investment and operating cost of bulk infrastructure of both vectors over representative days. The power system's operational constraints include ramping, unit commitment, state of charge for storage, and direct current (DC) power flow. Further details of the model are available in \cite{KhorramfarEtal-AE2024, KhorramfarEtal-NE2024}.

Our computational experiments are based on a 6-node power system and 23-node NG system representation of the New England region, considering projected 2050 energy demand, technology cost assumptions, and an 80\% decarbonization constraint with respect to 1990 levels \cite{Brattle2019}. Operational decisions and parameters in the power system exist at an hourly resolution while those in the NG system are at a daily resolution. The operational scenario set consists of $5$ training, $10$ validation, and $5$ test projections for $2050$. These stochastic scenarios contain parameters encoding both power-gas load and VRE capacity factors based on observations from $20$ weather-years after assuming moderate electrification of residential heating \cite{KhorramfarEtal-NE2024}.

We classify these parameters into five groups: (1) demand for electric power, (2) demand for NG, (3) solar CF, (4) onshore wind CF, and (5) offshore wind CF. Our BO approach tunes the relative weights of these parameter groups in the RPC distance function. Accordingly, we introduce hyperparameters $\theta_1,\dots,\theta_5$ and bound $\theta_i\in [0, 1] $ for all $i$. 
Since relative distances are only unique up to re-scaling of $\theta_1,\dots,\theta_5$, we impose that $\theta_1 + \dots + \theta_5 = 1$. We also introduce two hyperparameters specifying the number of non-extreme days, $5\leq\theta_6\leq80$, and the number of extreme days, $0\leq\theta_7\leq10$. Specifically, we introduce $\theta_7$ extreme days for both the power and NG systems according to total daily load so that $2\times\theta_7$ extreme days are included in total.

Using \textsc{BoTorch} \cite{botorch}, we implement two BO approachs with different GP hyperparameters, $\beta = 10$ and $\beta = 2$ (Eq.~\ref{eq:LCB}). We denote these two methods by $\text{BO}_{10}$ and $\text{BO}_{2}$ respectively.  We conduct four trials, each of which corresponds to an initialization with $N_0=20$ random function evaluations followed by 80 iterations of our BO-assisted approach. The resulting hyperparameters are evaluated according to their corresponding validation cost, $f^\text{val}$ (Sec.~\ref{sec:function_evaluation}), and test cost, $f^\text{test}$, defined to be the average cost incurred over the set of test projections. We also compare our approach to a random search heuristic that evaluates hyperparameters $\theta$ sampled uniformly at random from $\Theta$.

For all trials, we also report percentage improvement over two baseline solutions selected to represent the common practice in the literature of selecting a high number of representative days and clustering with all parameters weighted equally \cite{TeichgraeberBrandt2019, SunEtal2019}. In both baselines, we fix $\theta_1=\dots=\theta_5=0.2$ and set $\theta_6=80$. The two baselines are distinguished according to the number of extreme days they consider: the first (Base. 1) sets $\theta_7=0$ while the second (Base. 2) sets $\theta_7=10$. In our experiments, we also obtained results for a third baseline that uses $\theta_7=2$ extreme days and obtains the highest average validation and test costs among all results (\$14.38 and \$14.59 billion respectively). However, we omit this baseline from discussion of results due to space constraints.

\begin{table*}[htbp]
    \footnotesize
    \centering
    \caption{Hyperparameters, objective values, and improvement over baseline costs for each method (corresponding to lowest test cost across four trials). Percentage improvement is calculated with respect to the best performing baseline solution for the respective scenario set (i.e., Base. 1 for the validation set and Base. 2 for the test set). Dashes denote zero values.}
    \begin{tabular}{lccccccccccc}
        \toprule
        & \multicolumn{7}{c}{Aggregation Hyperparameters ($\theta^*$)} &  \multicolumn{2}{c}{Cost (\$ billion)} &  \multicolumn{2}{c}{Improv. (\%)} \\
        \cmidrule(lr){2-8} \cmidrule(lr){9-10} \cmidrule(lr){11-12}
        Method & Power & NG & Solar & Off. & On. & Rep. & Ext. & $f^\text{val}$ & $f^\text{test}$ & Val. & Test \\    
        \midrule
        Base. 1 & 0.2 & 0.2 & 0.2 & 0.2 & 0.2 & 80 & -- & 14.35 & 14.41 & -- & -0.29 \\
        
        Base. 2 & 0.2 & 0.2 & 0.2 & 0.2 & 0.2 & 80 & 10 & 14.42 & 14.37 & -0.5 & -- \\
        
        Random Search & 0.367 & 0.106 & 0.103 & 0.333 & 0.091 & 57 & -- & 13.97 & 13.93 & 2.7 & 3.0 \\
        
        BO ($\beta=10)$ & 0.389 & -- & 0.611 & -- & -- & 46 & -- & \textbf{13.87} & \textbf{13.82} & 3.3 & 3.8 \\
        
        BO ($\beta=2)$ & -- & 0.318 & 0.269 & 0.414 & -- & 64 & 2 & 13.93 & 13.89 & 2.9 & 3.4 \\
        \bottomrule
    \end{tabular}
    \label{tab:1}
\end{table*}

\begin{figure*}[htbp]
    \centering
    \begin{subfigure}[t]{0.35\textwidth}
        \centering
        \includegraphics[width=\textwidth]{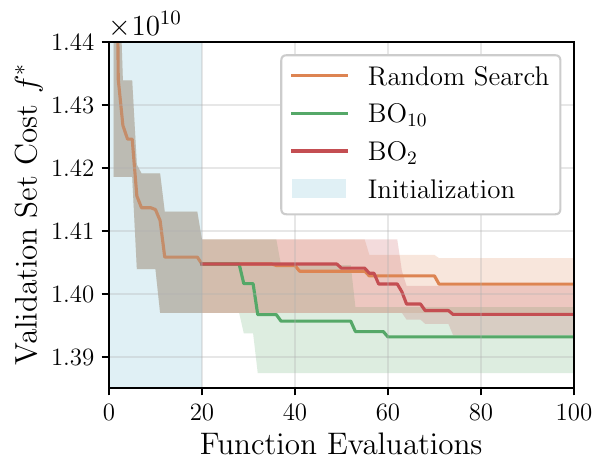}
        \caption{Lines show lowest validation costs obtained thus far averaged over all trials while shaded regions show highest and lowest costs obtained over all trials. `Initialization' denotes the random search phase used to estimate initial GP models.}
        \label{fig:raw_cluster_results}
    \end{subfigure}
    \hfill
    \begin{subfigure}[t]{0.61\textwidth}
        \centering
        \includegraphics[width=\textwidth]{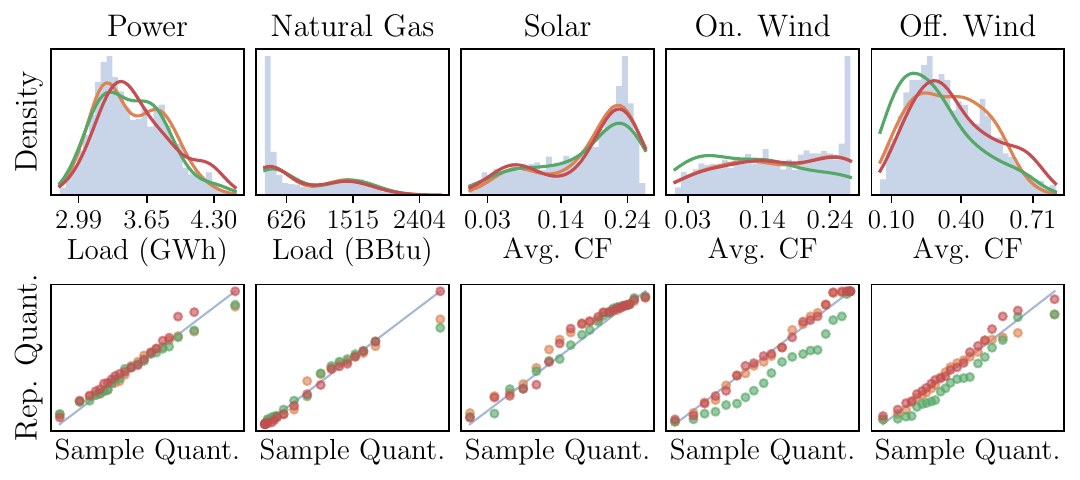}
        \caption{Top: histograms of daily parameter values observed in the sample set (blue) and in the rep. day sets corresponding to Table~\ref{tab:1} (shown as kernel density estimates). Values are averaged over hours and nodes for all parameters except gas load. Bottom: quantile-quantile plots comparing the empirical distribution of the sample set to the identified rep. day sets.}
        \label{fig:distributions}
    \end{subfigure}
    \caption{Clustering and objective cost results corresponding to the lowest-cost trial for each search method.}
\end{figure*}

\subsection{Cost Comparison}
Fig.~\ref{fig:raw_cluster_results} shows convergence of all three methods averaged over the four trials. 
The random search heuristic shows relatively slow convergence as it is not able to incorporate previous function evaluations to identify promising regions of the search space with effective capacity expansion decisions (or conversely, to avoid regions with ineffective capacity expansion decisions). On the other hand, $\text{BO}_{10}$ quickly converges to low-cost solutions. $\text{BO}_{2}$ also outperforms random search but converges more slowly and to higher-cost solutions. This might reflect the non-smooth nature of $f$ in small neighborhoods due to planning outcomes being sensitive to small changes to $\theta$. Consequently, an acquisition function that favors exploration (i.e., the one used for $\text{BO}_{10}$) outperforms one that evaluates multiple $\theta$ in a small neighborhood after obtaining one low-cost solution.

Table~\ref{tab:1} shows the best hyperparameter configuration identified for each method over all trials and its corresponding cost and improvement over the baselines. Interestingly, test costs are lower than validation costs across most experiments. This is a reflection of the relative harder-to-meet supply-demand projections that exist in the validation set rather than a statistical bias intrinsic to the optimization procedure. Importantly, all search methods improve over the baseline test costs by at least 3\% despite using fewer representative days. This demonstrates the necessity of tuning RPC hyperparameters over instantiating a large surrogate problem using generic (i.e., untuned) distance metrics.

\subsection{Hyperparameter Values}

The best hyperparameter configuration -- as evaluated by both validation and test costs -- is identified by $\text{BO}_{10}$. This configuration assigns relatively high weight to variability in solar availability and comparatively lower weight to offshore, and particularly onshore, wind availability. As a result, the distribution of daily power loads and solar CFs in the representative day set closely approximates the distribution observed in the larger sample set (Fig.~\ref{fig:distributions}). This is not the case for wind availability, which is largely underestimated by the representative day set. On the other hand, random search and $\text{BO}_{2}$ both learn to weight offshore wind CFs highly, and as a result, closely approximate the distribution of daily offshore wind availabilities observed in the sample set. Additionally, these methods also well-approximate the distribution of daily \textit{onshore} wind availabilities, which is reflective of the high correlation between onshore and offshore wind availabilities seen in Fig.~\ref{fig:corr}.

While we observe that generally at least one of either power or NG load is weighted highly, the total weight assigned to these parameters is less than that assigned to VRE CFs. This reflects the relative variability of VRE availability profiles, which are highly intermittent hour-to-hour, as compared to energy demand profiles. Moreover, demand for power is relatively correlated with demand for NG (Fig.~\ref{fig:corr}). As a result, a sufficient range of power and NG demand patterns will likely be captured as long as at least one of the two parameters is weighted highly during clustering (Fig.~\ref{fig:distributions}). This explains the weight of $0$ assigned to NG by $\text{BO}_{10}$, which nevertheless obtains low-cost planning outcomes.

The identified hyperparameter configurations also include at least 46 representative days but no more than 2 extreme days (if any). The absence of explicitly chosen extreme days in many of the instances indicates that the clustering approach sufficiently captures load volatility in the manner it partitions the planning horizon. In line with Fig.~\ref{fig:intro}, in which solutions were obtained from clustering a single scenario (with $\theta_1=\dots=\theta_5=0.2$), we observe that the selected number of representative days is less than the maximum number allowed. This again demonstrates that, in settings considering out-of-sample operational cost objectives, increasing the number of representative days does not necessarily lead to lower costs. On the other hand, the number of representative days chosen is greater than $25$, which yielded the lowest average out-of-sample cost for the case shown in Fig.~\ref{fig:intro}. This suggests that tuning distance weight hyperparameters and utilizing a larger sample set might help to reduce overfitting of planning outcomes to $\mathcal{T}_R$ when the number of representative days is high.

\subsection{New Assets and Utilization}
\begin{figure}[h]
    \centering
    \includegraphics[width=0.5\textwidth]{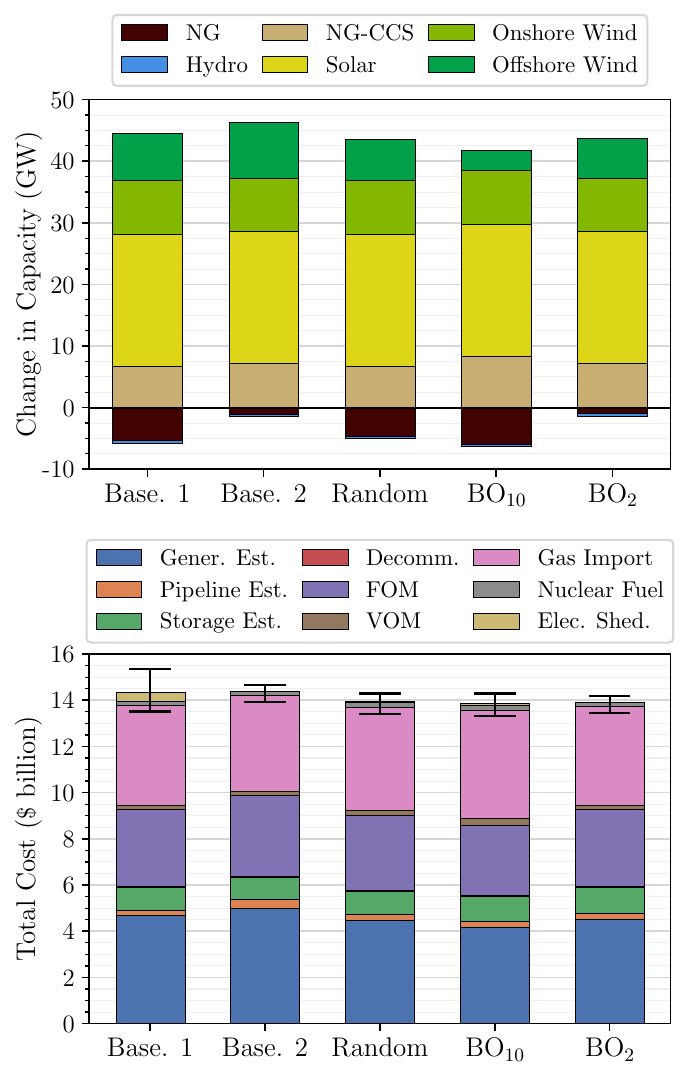}
    \caption{Capacity change for power plants (top) and cost breakdown (bottom). Error bars show maximum and minimum total cost incurred over the 15 projections that include validation and test sets. `NG' denotes existing gas-fired plants. `NG-CCS' is a gas-fired plant with carbon capture and storage technology. `FOM' and `VOM' are fixed and variable costs for plants.}
    \label{fig:investments}
\end{figure}

Fig.~\ref{fig:investments} shows installed capacity for power plants and cost components for all solved investment strategies. Comparing the two baselines, we find that including extreme days reduces costs from load shedding as well as inter-annual variations in operational costs (as shown by the narrower error bars for Base. 2). This is achieved by investing more aggressively in offshore wind generation while retaining a larger number of existing NG plants. Ultimately, this yields slightly higher costs on average, but significantly lower costs for worst-case projections as shown by the wider error bars for Base. 1 in Fig.~\ref{fig:investments}.

Of all methods, $\text{BO}_{10}$ yields the lowest average system cost. The resulting planning outcomes recommend high decommissioning and relatively low capacity addition for wind power (both onshore and offshore) compared to all other approaches. The extensive decommissioning of the existing gas plant is more than compensated by the investment in NG-CCS plants that operate on gas but capture most of the emissions \cite{KhorramfarEtal-AE2024}. This is consistent with the high import cost of gas in $\text{BO}_{10}$ in its cost breakdown. Relatively lower investment in offshore wind generation can be explained by underestimation of wind generation potential as captured by $\text{BO}_{10}$ (Fig.~\ref{fig:distributions}). As a result of high decommissioning and low generation expansion, the planning decisions identified by $\text{BO}_{10}$ yield higher costs for worst-case projections than $\text{BO}_{2}$ as shown by the wider error bars in Fig.~\ref{fig:distributions}. Ultimately, $\text{BO}_{10}$ only yields slightly lower expected costs than $\text{BO}_{2}$, and the decision to adopt either investment strategy should reflect the importance a planner places in worst-case costs.

\begin{table}[h]
    \footnotesize
    \centering
    \caption{Installed VRE capacity (GW), VRE utilization rate (\%), and storage investment (\$ billion). Utilization rate is calculated as average hourly generation divided by installed capacity.}
    \label{tab:2}
    \tabcolsep=0.11cm
    \begin{tabular}{lccccccc}
        \toprule
        & \multicolumn{2}{c}{Solar} &  \multicolumn{2}{c}{Offshore} &  \multicolumn{2}{c}{Onshore} & Storage \\
        \cmidrule(lr){2-3} \cmidrule(lr){4-5} \cmidrule(lr){6-7} \cmidrule(lr){8-8}
        & Cap. & Rate & Cap. & Rate & Cap. & Rate & Inv. \\
        \midrule
        Base. 1 & 21.4 & 2.71 & 7.7 & 6.00 & 8.7 & 8.48 & 1.00 \\
        Base. 2 & 21.4 & 2.66 & 9.0 & 5.98 & 8.7 & 8.49 & 0.96 \\
        Random & 21.4 & 2.74 & 6.6 & 6.06 & 8.7 & 8.48 & 1.02 \\
        $\text{BO}_{10}$ & 21.4 & 2.79 & 3.3 & 6.09 & 8.7 & 8.50 & 1.07 \\
        $\text{BO}_{2}$ & 21.4 & 2.76 & 6.5 & 6.06 & 8.7 & 8.51 & 1.12 \\
        \midrule
    \end{tabular}
\end{table}

Compared to Base. 1 and Base. 2, all search methods invest less in expanding generation capacity but more in energy storage infrastructure (Tab.~\ref{tab:2}). Nevertheless, their resulting investments can substantially reduce costs from load shedding in the power system when compared to Base. 1, and in the case of $\text{BO}_{2}$, operate with zero load shedding (similar to Base. 2) as a result of including two extreme days (Fig.~\ref{fig:investments}). Interestingly, Tab.~\ref{tab:2} shows that the utilization rate of VRE assets is generally higher for the search approaches than the baselines. In particular, $\text{BO}_{10}$, which weighted solar CFs more heavily in clustering, is able to dispatch more power from the same installed solar capacity through optimizing geographical distribution of plant locations. These findings reflect the cost savings obtained by our BO-assisted approach resulting from higher investment in storage coupled with establishment of relatively fewer VRE assets that can nevertheless be dispatched more effectively through strategic positioning.

\section{Concluding Remarks}
In this work, we presented a BO-assisted approach for solving large-scale stochastic capacity expansion problems that learns to construct and solve reduced-order models in deployment. Leveraging an established time series aggregation heuristic, we identify lower-cost capacity expansion decisions by optimizing over a continuous, low-dimensional space of representative period clustering hyperparameters. Importantly, we optimize with respect to operational costs incurred by planning decisions on a set of validation scenarios to obtain a less biased expected operational cost estimate. We apply our approach to capacity expansion planning of a coupled power and NG network and show that, when compared to conventional approaches to time series aggregation, our approach is able to reduce load shedding and other operational costs while minimizing investment costs through greater investment in storage and strategic positioning of solar and offshore wind generation assets.

In our discussion of results, we identified two investment strategies, both discovered by BO approaches, that yielded similar costs on average but different costs under worst-case projections. While we focus on minimizing expected costs in this work, future works may leverage recent advances in BO that consider risk measure-based objectives \cite{cakmak2020} to optimize for worst-case scenarios. Future work is needed to explore learning joint spatio-temporal aggregations of CEPs (following \cite{brenner2023}) to more directly consider the effects of geographical variation in VRE availability.

\section*{Acknowledgement}
The authors would like to acknowledge support from the MIT Energy Initiative Future Energy System Center and MIT Climate Grand Challenge ``Preparing for a new world of weather and climate extremes.''

\bibliographystyle{IEEEtran}
\bibliography{IEEEabrv, bibliography}

\end{document}

%% file: Preamble.tex
\usepackage[utf8]{inputenc}
\usepackage{amsfonts}
\usepackage{algorithm}
\usepackage{subcaption}
\usepackage{stfloats}
\usepackage{color,xcolor, url, float,  verbatim, changepage}
\usepackage{graphicx}
\usepackage{setspace}
\usepackage{array}
\usepackage{mathtools,amsmath,amssymb} 
\usepackage{hyperref}
\usepackage{lmodern}
\usepackage[T1]{fontenc}
\usepackage{textcomp}
\usepackage{booktabs}
\usepackage{babel} 
\usepackage{algpseudocode}
\usepackage[]{geometry}

\usepackage{amsthm}

\usepackage{xcolor}
\usepackage{soul}

\definecolor{blue}{rgb}{0.38, 0.51, 0.71} 
\definecolor{darkblue}{RGB}{17, 42, 60} 
\definecolor{red}{RGB}{175, 49, 39} 

\definecolor{orange}{RGB}{217, 156, 55} 
\definecolor{green}{RGB}{144, 169, 84} 
\definecolor{palegreen}{RGB}{197, 184, 104} 

\definecolor{yellow}{RGB}{250, 199, 100} 
\definecolor{brokenwhite}{RGB}{218, 192, 166} 
\definecolor{brokengrey}{rgb}{0.77, 0.76, 0.82} 

%% file: tikz_set.tex
\tikzstyle{block} = [draw, fill=white, rectangle, 
    minimum height=3em, minimum width=6em]
\tikzstyle{sum} = [draw, fill=white, circle, node distance=1cm]
\tikzstyle{input} = [coordinate]
\tikzstyle{output} = [coordinate]
\tikzstyle{pinstyle} = [pin edge={to-,thin,black}]
\makeatletter
\tikzset{
    database top segment style/.style={draw},
    database middle segment style/.style={draw},
    database bottom segment style/.style={draw},
    database/.style={
        path picture={
            \path [database bottom segment style]
                (-\db@r,-0.5*\db@sh) 
                -- ++(0,-1*\db@sh) 
                arc [start angle=180, end angle=360,
                    x radius=\db@r, y radius=\db@ar*\db@r]
                -- ++(0,1*\db@sh)
                arc [start angle=360, end angle=180,
                    x radius=\db@r, y radius=\db@ar*\db@r];
            \path [database middle segment style]
                (-\db@r,0.5*\db@sh) 
                -- ++(0,-1*\db@sh) 
                arc [start angle=180, end angle=360,
                    x radius=\db@r, y radius=\db@ar*\db@r]
                -- ++(0,1*\db@sh)
                arc [start angle=360, end angle=180,
                    x radius=\db@r, y radius=\db@ar*\db@r];
            \path [database top segment style]
                (-\db@r,1.5*\db@sh) 
                -- ++(0,-1*\db@sh) 
                arc [start angle=180, end angle=360,
                    x radius=\db@r, y radius=\db@ar*\db@r]
                -- ++(0,1*\db@sh)
                arc [start angle=360, end angle=180,
                    x radius=\db@r, y radius=\db@ar*\db@r];
            \path [database top segment style]
                (0, 1.5*\db@sh) circle [x radius=\db@r, y radius=\db@ar*\db@r];
        },
        minimum width=2*\db@r + \pgflinewidth,
        minimum height=3*\db@sh + 2*\db@ar*\db@r + \pgflinewidth,
    },
    database segment height/.store in=\db@sh,
    database radius/.store in=\db@r,
    database aspect ratio/.store in=\db@ar,
    database segment height=0.1cm,
    database radius=0.25cm,
    database aspect ratio=0.35,
    database top segment/.style={
        database top segment style/.append style={#1}},
    database middle segment/.style={
        database middle segment style/.append style={#1}},
    database bottom segment/.style={
        database bottom segment style/.append style={#1}}
}
\makeatother

%% file: Tikz_Figures/learning-to-configure.tex
\node [database, name=input, database radius = 0.4cm, database segment height=0.2cm, database middle segment={fill=red!10}, database bottom segment={fill=red!10}, database top segment={fill=red!10}] (input) {};
\node [align=center] (dataset) {\\\\\\\\\\Instance\\Dataset};

\node [block, right of=input, align=center, node distance=3.2cm, fill=blue!10] (controller) {Heuristic};
\node [block, right of=controller, node distance=3cm, align=center, fill=green!10] (objective) {Objective\\Value};
\node [below of=controller, align=center] (theta) {$\theta$};

\draw [->] (controller) -- node[align=center, name=u, above] {$x$} (objective);
\node [output, right of=objective] (output) {};
\node [draw, diamond, below of=u, align=center, fill=orange!10] (ML) {ML};

\draw [->] (input) -- node [align=center] {Instance} (controller);
\draw [dashed, ->] (objective) |- (ML);
\draw [->] (theta) -- (controller);

%% file: Tikz_Figures/learning-to-configure-deploy.tex
\node [input, name=input] (input) {};
\node [block, right of=input, align=center, node distance=2.7cm, fill=blue!10] (controller) {Heuristic};
\node [right of=controller, node distance=2cm] (objective) {};

\draw [->] (controller) -- node[align=center, name=u, above] {$x$} (objective);
\node [output, right of=objective] (output) {};
\node [draw, diamond, below of=u, align=center, fill=orange!10] (ML) {ML};

\draw [->] (input) -- node [align=center] {Instance} (controller);
\draw [->] (ML) -| node {} 
    node [near end] {$\theta$} (controller);

%% file: Tikz_Figures/BO.tex
\node [input, name=input] (input) {};
\node [block, right of=input, align=center, node distance=2.7cm, fill=blue!10] (controller) {Heuristic};
\node [block, right of=controller, node distance=3cm, align=center, fill=green!10] (objective) {Objective\\Value};

\draw [->] (controller) -- node[align=center, name=u, above] {$x$} (objective);
\node [output, right of=objective] (output) {};
\node [draw, diamond, below of=u, align=center, fill=orange!10] (ML) {ML};

\draw [->] (input) -- node [align=center] {Instance} (controller);
\draw [dashed, ->] (objective) |- (ML);
\draw [->] (ML) -| node {} 
    node [near end] {$\theta$} (controller);